\documentclass[10 pt]{article}
\usepackage{epsfig}
\begin{document}

\title{\bf Faster than Light Quantum Communication}

\author{A.Y. Shiekh\footnote{\rm shiekh@dinecollege.edu} \\
             {\it Din\'{e} College, Tsaile, Arizona, U.S.A.}}

\date{}

\maketitle

\abstract{Faster than light communication might be possible using the collapse of the quantum wave-function without any accompanying paradoxes.}

\baselineskip .5 cm

\section{Introduction}

It has long been wondered if faster than light communication might be possible~\cite{Fayngold} and the collapse of the quantum wave-function, upon measurement, might be utilized to achieve this, with due concern for any paradoxes that might result.

Firstly, there is no direct violation of special relativity since it is the quantum wave-function that collapses and no energy or matter travels at faster than light speed.

\section{Unitary communicator}
The conservation of a particle in quantum theory (unitarity) might suggest a possible mechanism, since the destructive interference in one part of the system will imply a greater probability of locating the particle in another part, no matter how dispersed the system has become.\footnote{A similar mechanism has been proposed to augment the ability of a quantum computer~\cite{Shiekh1},~\cite{Shiekh2}.}

To try and implement this, imagine a beam splitting mechanism that breaks the beam into two arms that can be widely separated, and then again splits and recombines one of the two resulting arms.

\begin{center}
\includegraphics[scale=.35]{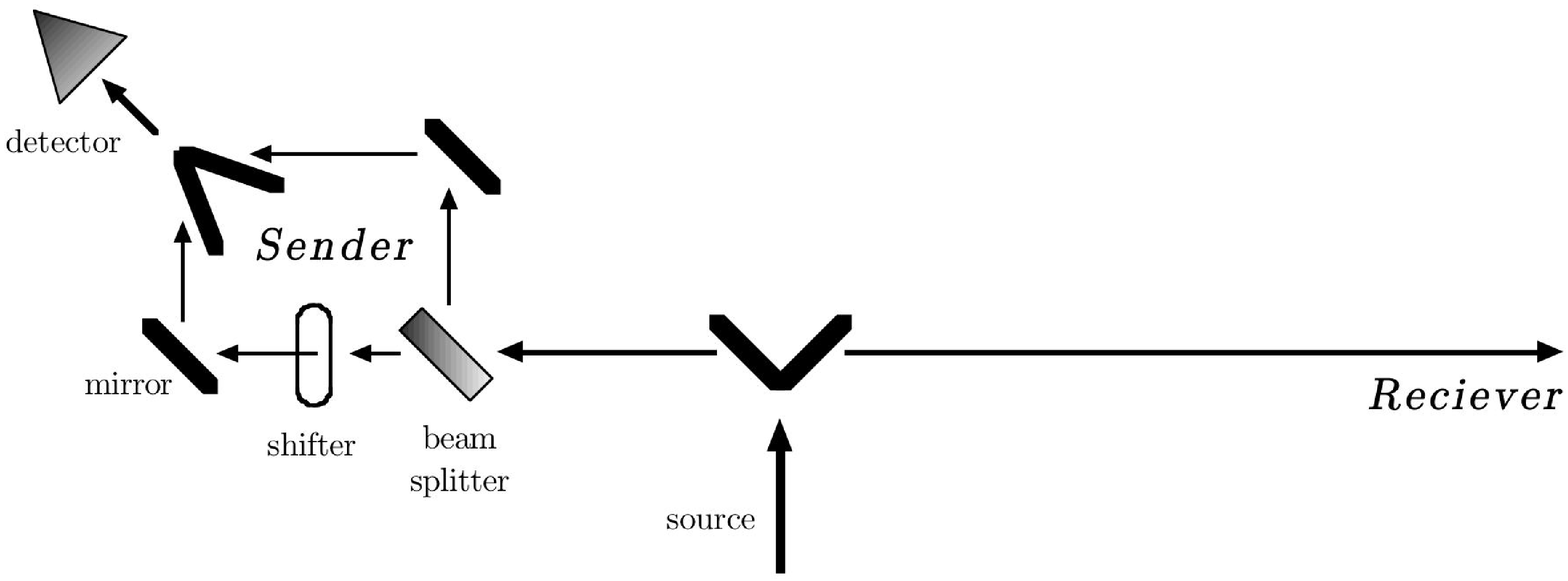}
\end{center}
\begin{center}
Quantum Transmitter
\end{center}

The recombination can be arranged to constructively, or destructively interfere, depending on a phase shifter in one of the two paths.

If the sender arranges for constructive interference then some of the particles will be `taken up' by the sender, but none if destructive interference is arranged; in this way the intensity of the receivers beam might be controlled. So a faster than light transmitter of information (but not energy or matter) might be possible.

\section{Two-way communication}
The above proposal, for simplicity, was a one way transmission device, but this can be easily duplicated for a full-duplex device or simply extended for half-duplex.

\begin{center}
\includegraphics[scale=.35]{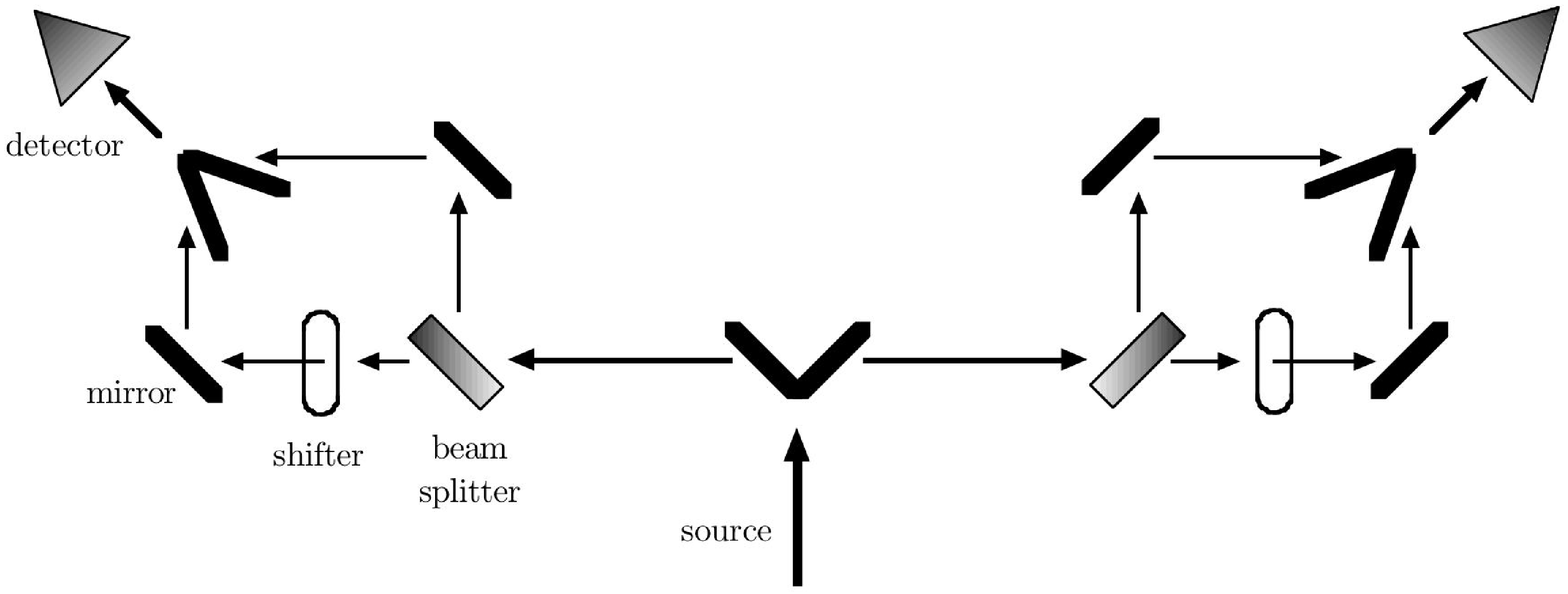}
\end{center}
\begin{center}
Half-Duplex Quantum Communicator
\end{center}

\section{Absence of Paradox}
The above proposal would seem more plausible if it can be demonstrated that no paradox arises from its supposed ability to communicate `instantly' over indefinite distances; namely, that no use can be made of a communication to alter events in the past.

The collapse of the wave-function upon the act of measurement has long been a dilemma \cite{Scarani, Ghirardi1}, and one seeks to explain when and how the reduction occurs, if at all.

A possible clarification to the usual quantum measurement axiom might be that the probabilistic collapse happens when distinguishability occurs, and that `instantaneous' might make more sense if relative to a preferred frame (a quantum-ether).

\subsection{When does the collapse occur}
Taking as the model of a `macroscopic' system, the interference of large molecules, which has been performed \cite{Constance}; if the energy of an outer electron is modified on one path alone (which is not enough to wash out the interference pattern), interference is still lost. This is due to distinguishability, as one would then know which path the object took from the observed state of the outer electron.

If one now goes a step further and argues that distinguishability not only stops interference, but actually triggers the collapse of the wave function (is itself the act of measurement), one may have another view on the question of where the boundary between the quantum and classical worlds occurs. This should not lead to any new prediction, since the quantum effect (interference) is no longer present anyhow.

Further, since the act of distinguishability inevitably involves the interaction with another system, there is no dilemma with momentum/energy non-conservation, as would be the case in proposals invoking spontaneous reduction, such as the GRW model.

\subsection{How fast does the collapse happen}
It is said that the collapse of the wave-function happens `instantly', but as is well known, relativity does not respect this concept; what is instant in one frame is not in another. It also does not seem reasonable that a moving measuring device would instigate a different collapse from a stationary one, and one way around this dilemma is that there is a preferred frame in which the collapse occurs.

Up until now this was not a pressing issue for, although it would alter the cause and effect ordering for the measuring of an EPR pair, the end result was not influenced by which end made the measurement first. This uneasy state of affairs is brought to a head here, but fortunately the above suggestion that the collapse occurs in some preferred frame also severs to prevent the faster than light proposal from being able to communicate into the past.

Other backward time travel proposals, such as worm holes, have been refuted~\cite{Shiekh3}.

\section{Conclusion}
Faster than light communication may be possible using the collapse of the wave-function, and without any paradoxical powers accompanying the device.

These proposals for when and how the measurement occurs might clarify, in a natural way, why simple systems  such as elementary particles express their quantum nature so easily, and why more structured systems do not. It is not necessary to explain why the hypothesized mechanisms operate, anymore than Newtonian gravity explains why a mass exerts a force or Einstein gravity sees it as warping space-time.

\end{document}